\documentclass[12pt]{article}
\usepackage[pctex32]{graphics}
\begin{document}
\begin{center}
  
{\Large \bf Boson-Fermion Transmutation and the Statistics of Anyons}

\vspace{1cm}

                      Wung-Hong Huang*\\
                       Department of Physics\\
                       National Cheng Kung University\\
                       Tainan,70101,Taiwan\\

\end{center}
\vspace{2cm}

\begin{center} {\large \bf ABSTRACT }\end{center}

   It is shown that, by allowing a transmutation between a boson and a fermion, the system with both bosons and fermions will have the statistical distribution function of an anyon.

\vspace{4cm}
\begin{flushleft}
*E-mail:  whhwung@mail.ncku.edu.tw\\
Physical Review E51 (1995) 3729
\end{flushleft}


\newpage
In recent years, "anyons," which are considered to  obey "fractional" statistics [1,2], have been a subject of  intense study. The anyons have found application in the  fractional quantized Hall effect [3] and high temperature superconductor [4]. Most of the study has been done in the context of many-body quantum mechanics and, in a recent paper [5], Wu derived the occupation-number distribution function of the anyon gas to formulate the theory of quantum statistical mechanics.  Using the idea of fractional exclusion [6], the statistical weight function $W_{anyon}$ for the system with $N$ identical anyons occupying
a group of G states is given by

        $$W_{anyon} = {[ G + (N - 1 )( 1 - \alpha ) ]!\over  N![G-\alpha N-(1-\alpha)]!},          \eqno{(1)}$$
\\
where $\alpha$ is the fraction statistical parameter and $\alpha =0$, corresponding to bosons, while $\alpha =1$ fermions. Then, through the standard manipulation [7], one can obtain the most probable value of the statistical distribution function of an anyon, which is derived by Wu [5].

   In this Brief Report we will show that, for a system with the $\alpha$ fraction of fermion and the $(1-\alpha )$ fraction of boson, then, once we allow the transmutation between the boson and fermion, the system will have the statistical
distribution function, just like that of an anyon. The derivation is very simple. As usual, the statistical weight functions for the systems with $\alpha N$ fermions or $(1-\alpha  N)$ bosons occupying a group of G states are given by

$$W_b = {[ G-1 + ( 1 - \alpha )N ]!\over  [(1-\alpha) N]!(G-1)!},~~~~ W_f = { G!\over  (\alpha N)!(G -\alpha N)!}.          \eqno{(2)}$$

   Once we allow the transmutation between the boson and the fermion, then it is meaningless to justify which particle is a boson or a fermion, even though the particle that is a fermion is still required to obey the exclusion principle. Thus, the statistical weight function $W$ becomes

$$W_{b+f} = W_b W_f \left [{N! \over  [( 1 - \alpha )N ]! \alpha N!}\right ]^{-1} = {[ G -1 + ( 1 - \alpha )N ]!G!\over (G-1)! [G - \alpha N]!}. \eqno{(3)}$$

   In the thermodynamical limit, as both G and N are very large, the statistical weight function $W_{b+f}$ is equal to $W_{anyon}$.  Thus both systems have the same statistical distribution functions. This completes our proof.

   It shall be mentioned that there is a literature [8] that also investigates the property of transmutation from a state in which the particle obeys a statistics to another state in which the particle obeys another statistics. However, the author of Ref.[8] mainly discusses the classical model of intermediate statistics in which the particles are Brownian and obey the classical kinetic equation. After
analyzing the classical nonlinear kinetics of Brownian particles and taking into account the exclusion and inclusion principle, the distribution function obtained in Ref. [8] does not become the anyon distribution. On the other hand, in our analyses there are already two kinds of quantum particles (boson and fermion) in the beginning, and we allow the transmutation between these two kinds
of quantum particles.

   It is hoped that the property of the equivalence between the anyon statistics and statistics with boson-fermion transmutation that we have found will be helpful in explaining or predicting the statistical property of anyons.

\newpage
{\begin{center} {\large \bf  REFERENCES} \end{center}}
\begin{enumerate}

\item  J. M. Leinaas and J. Myrheim, Nuovo Cimento B 38, 1
   (1977).
\item F. Wilczek, Phys. Rev. Lett. 48, 1144 (1982); 49, 957
   (1982).
\item  R. B. Laughlin, Phys. Rev. Lett. 50, 1395 (1983); Phys. Rev. B 27, 3383 (1983);\\
       B. I. Halperin, Phys. Rev. Lett. 52, 1583 (1984); 52, 2390(E) (1984);\\
       R. Prange and S. M. Girvin, The Quantum Hall Effect (Springer, Berlin, 1987).
\item  P. B. Wiegmann, Phys. Rev. Lett. 60, 821 (1988); \\
       R. B. Laughlin, ibid. 60, 2677 (1988); \\
      Y.-H. Chen, F. Wilczek, E. Witten, and B. I. Halperin, Int. J. Mod. Phys. B 3, I001 (1989).
\item  Y.-S. Wu, Phys. Rev. Lett. 73, 922 (1994).
\item  F. D. M. Haldane, Phys. Rev. Lett. 67, 937 (1991);\\
      M. D.  Johnson and G. S. Canright, Phys. Rev. B 49, 2947 (1994);\\
   M. V. N. Murthy and R. Shankar, Phys. Rev. Lett. 72, 3629 (1994).
\item  See, e.g., K. Huang, Statistical Mechanics (Wiley, New
   York, 1963).
\item  G. Kaniadakis, Phys. Rev. E 49, 5111 (1994).

\end{enumerate}
\end{document}